\documentclass[letterpaper,twocolumn,prl,showpacs]{revtex4-1}
\usepackage[latin9]{inputenc}
\usepackage{amsmath}
\usepackage{amssymb}
\usepackage{graphicx}
\usepackage{esint}

\makeatletter

\pdfpageheight\paperheight
\pdfpagewidth\paperwidth

\@ifundefined{textcolor}{}
{%
 \definecolor{BLACK}{gray}{0}
 \definecolor{WHITE}{gray}{1}
 \definecolor{RED}{rgb}{1,0,0}
 \definecolor{GREEN}{rgb}{0,1,0}
 \definecolor{BLUE}{rgb}{0,0,1}
 \definecolor{CYAN}{cmyk}{1,0,0,0}
 \definecolor{MAGENTA}{cmyk}{0,1,0,0}
 \definecolor{YELLOW}{cmyk}{0,0,1,0}
}

\usepackage{epsfig}\usepackage{color}\usepackage{booktabs}

\makeatother

\begin{document}

\title{Record Dynamics: Direct Experimental Evidence from Jammed Colloids}

\author{Dominic M. Robe$^{1}$, Stefan Boettcher$^{1}$, Paolo Sibani$^{2}$,
and Peter Yunker$^{3}$}

\affiliation{$^{1}$Department of Physics, Emory University, Atlanta, GA 30322,
USA }
\affiliation{$^{2}$FKF, Syddansk Universitet, DK5230 Odense M, Denmark}
\affiliation{$^{3}$School of Physics, Georgia Institute of Technology, Atlanta,
GA 30332, USA }
\begin{abstract}
In  a broad class of  complex materials a quench leads  to a multi-scaled  relaxation process known as aging.
To explain its commonality and the astounding insensitivity to most microscopic details,
record dynamics (RD)  posits that  a small set of  increasingly rare and irreversible
events, so called \emph{quakes}, 
controls the dynamics. While key  predictions  of RD are known  to  concur with a number of  experimental 
and simulational results, its basic assumption on the nature of quake statistics 
 has proven extremely difficult to verify experimentally. The careful distinction of  rare ("record") 
cage-breaking events from  in-cage rattle accomplished in previous 
experiments on jammed colloids, enables us to extract 
 the first \emph{direct} experimental evidence for the fundamental hypothesis of RD that
 the rate of quakes  decelerates with the inverse of the  system age.
The resulting description  shows the  predicted  growth of the particle  mean square displacement and of 
a mesoscopic  length-scale with the logarithm of time.
\end{abstract}


\maketitle
\section{Introduction}
 \vspace{-0.25cm}
Aging is a   decelerating   process
present  in manifold complex materials relaxing after a quench. It has  
 been studied 
for  decades  from different  
perspectives~\cite{Buisson03,Buisson03a,Bissig03,Vollmayr04,Yunker09,Kajiya13,Zargar13,Sibani05,Crisanti04,Sibani06b,Sibani06a,Oliveira05,Zotev03,Rodriguez03,Rodriguez13,Richardson10,Parsaeian08,Fernandez13,Vollmayr-Lee16,Courtland03,BoSi09,Becker14},
with the experimental focus moving, over the years,  from  relations  
such as the Fluctuation-Dissipation theorem and its violation~\cite{Buisson03,Buisson03a} on 
to the discovery, observation and study of the `anomalous'  non-equilibrium 
events~\cite{Bissig03,Vollmayr04,Yunker09,Kajiya13,Zargar13,Fernandez13} now
 recognized  as   key  properties. Concomitantly, theoretical and numerical studies have   
also considered  anomalous events~\cite{Crisanti04,Sibani05,Sibani06b}, and spin-glass thermo-remanent
 magnetization~\cite{Sibani06a}, magnetic flux creep in type-II high-$T_c$ superconductors~\cite{Oliveira05}, 
 ants moving out of their nest~\cite{Richardson10} and particle motion  in dense 
colloids~\cite{Courtland03,BoSi09,Becker14} have been interpreted using  the statistics  of
`quakes'. These  increasingly  rare cooperative changes are
spatially localized within domains or clusters and lead the system from one  metastable states to the next. 
The  coarse-graining   
scheme referred to as `record dynamics',
 links quakes with crossings of  
record high peaks in the  free-energy landscape~\cite{Sibani89,Sibani13} associated to 
 growing clusters and generated by the well-known exponential proliferation of local minima (`inherent states') with growing energy~\cite{Stillinger84,Stillinger99}. In dense colloidal 
suspensions,   quakes  are usually  called `cage breakings' while
reversible quasi-equilibrium (or 
Gaussian) fluctuations
are called 
`cage rattlings'~\cite{Weeks00,Yunker09,Hunter12}. 

Record dynamics has 
been proposed  as a tool to model complex systems~\cite{Anderson04,Sibani13} based on  its ability to 
explain in a unified fashion the  phenomenology of many different aging systems   
 by  filtering  away all their microscopic peculiarities. 
 The concurring  experimental evidence pertains  to macroscopic properties of 
  magnetic glasses~\cite{Crisanti04,Sibani06b,Sibani06a} and  dense colloidal 
suspensions~\cite{Courtland03,BoSi09}.
 However, direct experimental  support of  its basic microscopic assumption is  hard to
 gain and has been unavailable so far.

\begin{figure}
 \hfill{}\includegraphics[bb=0bp 0bp 740bp 530bp,clip,width=0.9\columnwidth]{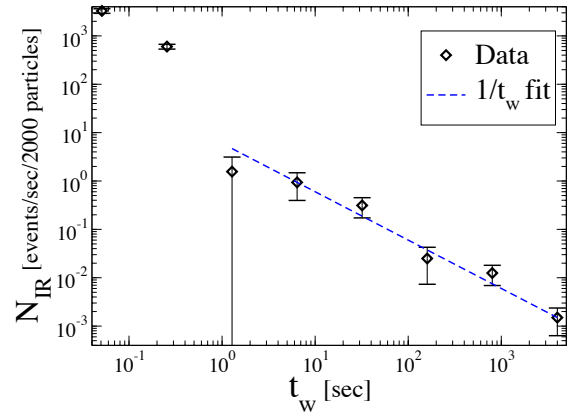}\hfill{}
 \vspace{-0.3cm}
\protect\caption{\label{fig:Decay}
Decay in the rate of intermittent cage-breaks (``quakes'') in a dense colloid. Re-plot on log-scale of the 
experimental data in Fig.~2a of Yunker et al.~\cite{Yunker09}. The data directly validates the $1/t$-decay 
fundamental to record dynamics that justifies the log-Poisson statistics~\cite{BoSi09,Becker14}.}
 \vspace{-0.3cm}
\end{figure}

\begin{figure}
 \vspace{-0.1cm}
 \hfill{}\includegraphics[bb=0bp 250bp 800bp 580bp,clip,width=0.9\columnwidth]{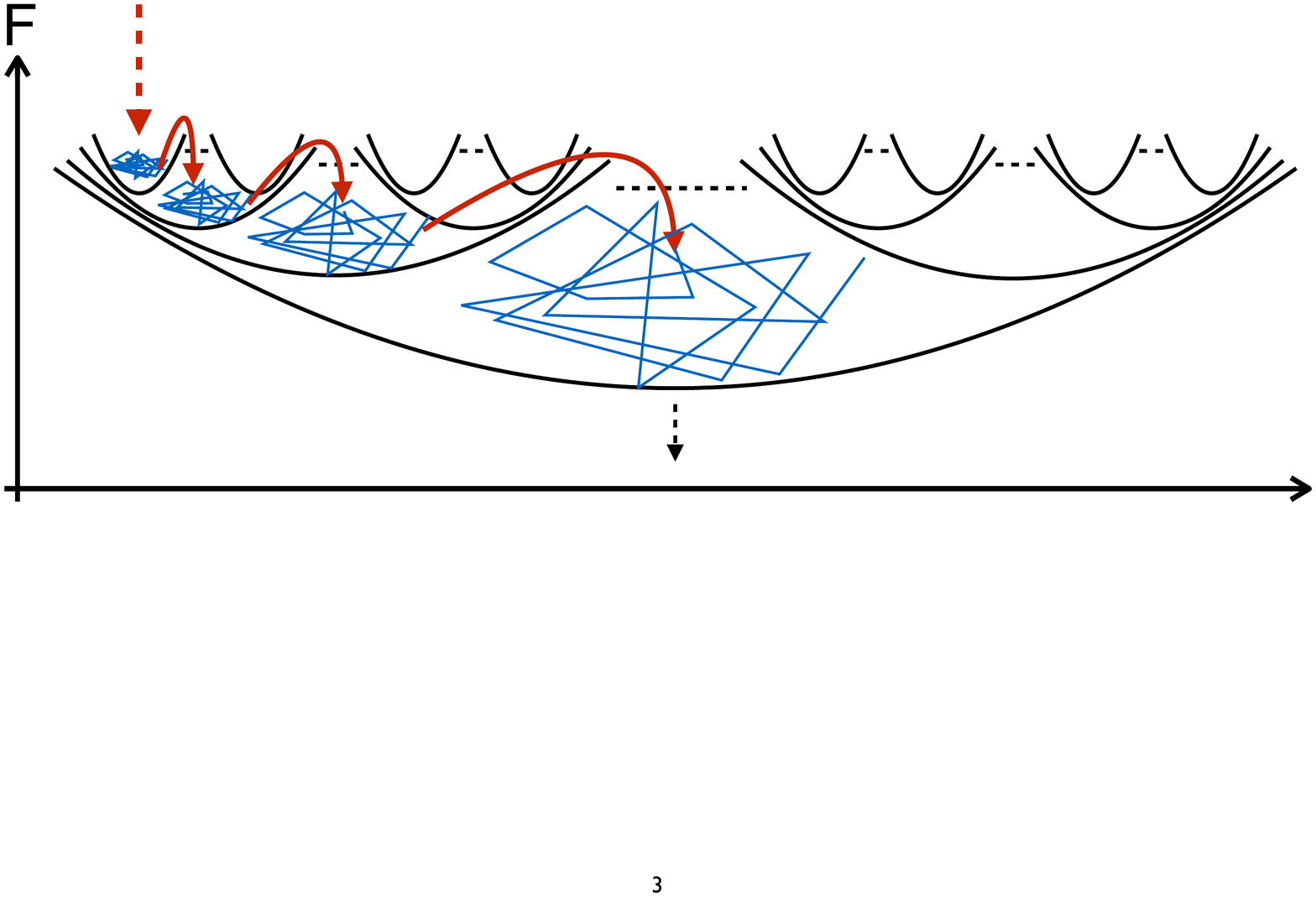}\hfill{}
 \vspace{-0.2cm}
\protect\caption{\label{fig:FLandscape}
Schematic view of the  hierarchical free-energy landscape of a small  jammed system (solid black), with a 
typical trajectory of an aging dynamics (blue and red). With increasing free energy $F$, local 
minima proliferate exponentially but also become shallower.  The black-dashed 
down-arrow signifies that the hierarchy continues to a distant ground state. 
The  dynamics generically 
evolves after a quench (red dashed arrow) through a sequence of quasi-equilibrium explorations 
(blue trajectories) and rare, intermittent quakes over record barriers (red arrows) that 
access an exponentially expanding portion of the configuration space.}
 \vspace{-0.3cm}
\end{figure}

The  data analyzed here stem from  three experimental runs by  Yunker et al~\cite{Yunker09} who
employed an aqueous bi-disperse suspension of micro-gel 
spheres sandwiched between glass cover slips to form a quasi-2D colloidal system with the
unique property that 
a temperature change of  $\approx 4$K at 
 room temperature,  could change the area fraction by $\approx10\%$ 
in $\approx0.1$s.
The systems were prepared at a high packing 
fraction, heated uniformly by a mercury lamp to form a colloidal fluid, then quenched to form a glass by removing 
the light. Final area fractions studied were in the range 81-84$\%$. A window of 2500 particles, embedded in a sea 
of  300000, was then tracked for $10^4$s.
Most of the data presented here and  in Ref.~\cite{Yunker09} stem  from  samples at the highest packing fractions
and exhibit the strongest aging behavior, while our analysis of the loosest 
packings  shows deviations  from it. In  Ref.~\cite{Yunker09}, dynamic heterogeneity and aging 
were studied through analysis of changing particle neighborhoods. Here we 
use  a threshold of $0.04\mu m$ to
separate small  reversible in-cage motion from irreversible  displacements.
 Importantly, once
irreversible configurational changes  are identified, the 
significance of rare  individual cage breaks for intermittency is demonstrated, and  the connection  
between small and  seemingly  insignificant localized dislocations and much larger collective shifts in 
particle position is established. 

These experiments provide the first  opportunity to 
confront both the  central assumption of record dynamics on  quake statistics, as shown in Fig.~\ref{fig:Decay}, and
one of its predictions, namely the emergence of growing mesoscopic real-space structures.
In this Letter, we  analyze the data of Ref.~\cite{Yunker09} to
show  that they concur with the fundamental assumptions as well as with the  predictions of the theory. 
For the first time, direct experimental evidence for record dynamics as a coarse-grained description of dense 
colloids in particular, and of a broad class of aging materials in general, is provided.

\section{Theoretical Background}
 \vspace{-0.25cm}
The relevance of records is 
immediate in the ``Backgammon model''~\cite{Ritort95}, in which $m$ particles are distributed over 
$n$ boxes ($m\!\gg\! n\!\gg\!1$) and where each update swaps one particle  randomly between boxes. A 
box which empties out by a chance fluctuation of size $\sim m/n$ becomes inaccessible as the 
system's energy corresponds to the number of occupied boxes. The dynamics thus requires a sequence 
of \emph{rare} record fluctuations of \emph{marginally} increasing average size $\sim m/n$, $\sim 
m/(n-1)$, $\sim m/(n-2), \dots$, to progress ever more slowly towards its ground state where all particles 
fit into \emph{one} box. The importance of  intermittent, irreversible record-events that are 
decorrelated by an exponential separation of time-scales is obvious. In Ref.~\cite{Becker14}, a simple 
lattice model was introduced where mobile particles accrete into jammed clusters only to be re-mobilized after a 
chance fluctuation at a time exponential in the size of the cluster. Following a quench at $t=0$ when all 
particles are mobile, clusters form and break up ever more slowly because every break-up removes  
 one cluster and  increases the average size of the remaining clusters. We will 
use simulations of this model for comparisons with the experimental data.
Finally, the connection between decelerating dynamics and growing mesoscopic real-space objects, our clusters 
or domains,  is studied theoretically in Ref.~\cite{Sibani16}, using a simple kinetically constrained 
model, the "parking lot model"~\cite{Krapivsky94b}, which has been applied in Ref.~\cite{Nowak98} 
to explain the logarithmically slow compactification of  granular piles.

In disordered systems with a large number of degrees of freedom, the concept of ``marginal 
attractor stability'' first introduced in~\cite{Tang87} is of central importance. In our context it 
presupposes three features that are well-established for the coarse-grained free-energy landscapes of 
complex systems, as illustrated in Fig.~\ref{fig:FLandscape}: (1) Meta-stable states with their 
combined basin of attraction, proliferate exponentially for increasing free energy~\cite{Bray81,Stillinger84,Stillinger99,Heuer08}; 
(2) More-stable (lower-energy) states typically have higher barriers to escape their 
basin~\cite{Sibani93b,Heuer08}; and (3) overcoming increasingly higher barriers makes exponentially more 
configurations accessible~\cite{BoSi}. The ensuing dynamics is also illustrated in Fig.~\ref{fig:FLandscape}:  In a 
quench, the system almost certainly gets stuck initially in a shallow basin of 
high energy. There,  a small, random fluctuation already suffices to escape into  a larger basin containing  
many sub-basins, some of which feature local minima  of lower energy. But for basins of lower energy, 
gradually higher fluctuations are required to escape. The  gain in stability acquired in any one of these 
escapes will most probably only be \emph{marginal}, since spontaneously arriving at minima of 
dramatically lower energy would be exponentially unlikely. In such a marginally deeper basin, the 
motion in and out of  states less stable than the original basin only provides reversible quasi-equilibrium  
fluctuations.  For an irreversible quake, typically, merely a record fluctuation in the random noise impinging on the system is 
required~\cite{Sibani93a}. In a spin glass a record-high  thermal energy 
fluctuation  elicits  a quake associated with  a large decrease in energy~\cite{Sibani05}. In  a colloidal glass, record-sized
fluctuations of locally available free space are needed to accommodate a new particle, and the final effect 
of a quake  is a density increase.

In this manner, on exponentially longer timescales, ever larger and rare fluctuations become possible 
such that incrementally higher barriers can be scaled in a sequence of intermittent record-sized events, 
thereby making an exponentially growing number of configurations accessible. This  approximation 
supersedes the particular  properties of  the  hierarchies of barriers in a given system.
 As long as such a hierarchy 
exists, i.e.,  the system is actually jammed, these differences only vary the overall unit of time. To 
summarize, if barrier heights only grow marginally  from one quake to the next, quake statistics can be 
meaningfully approximated by the statistics of  record sized fluctuations  in  the random white noise 
which drives the dynamics~\cite{Sibani93a}. Record dynamics thereby offers an analytical in-road to 
coarse-grained descriptions able to straddle the aging phenomenology.  

\section{Results from Experimental Data}
\label{data}
 \vspace{-0.25cm}
\subsection{Statistics of Record-Sized Events}
 \vspace{-0.25cm}
 The first 
key property of the experimental data  of Ref.~\cite{Yunker09}  that we check  is that irreversible quakes are generated at a decelerating rate  $\lambda(t)
\propto1/t$, like records in any \emph{iid} sequence of independent random numbers. Such events 
were identified as movements of particles involving a replacement of at least three nearest neighbors, 
shown in Fig.~2a of  Ref.~\cite{Yunker09}. In our Fig.~\ref{fig:Decay}, we bin the same data 
logarithmically and see that, over more than three decades, the rate of irreversible events  decelerates as 
$1/t_{\rm w}$,  with time $t_{\rm w}$ elapsed after the quench. Only an initial period of $\sim0.1$s, 
consistent with  the time needed for the quench, shows significant deviations from the record-dynamics 
prediction. To obtain the  data, Yunker et al.~\cite{Yunker09} had  to filter out  the ``in-cage rattle'', 
apparent from the inset of their Fig.~2a. This point is also emphasized by their  Fig.~2b, that shows that  
a slow but steady increase of the domain-size of correlated events can be detected if  reversible 
fluctuations are removed. Unfortunately, the sub-linear but perceptible growth in the relevant regime, 
i.e., for times $>0.1$s, can not conclusively shown to be logarithmic with time, as record dynamics 
would predict~\cite{BoSi09,Becker14}.

\subsection{Mean-Square Displacement}
 \vspace{-0.25cm}
As significant particle motion is activated by these quake events, the distance traveled is proportional to the integral 
of the rate $\lambda(t)\sim1/t$. Thus, the mean square displacement (MSD) of  particles moving between times
$t_{\rm w}$ and $t\ge t_{\rm w}$ in a  dense colloid is predicted by record dynamics~\cite{BoSi09,Sibani13} 
to grow as $\langle \Delta x^2(t,t_{\rm w}) \rangle\propto \ln (t/t_{\rm w})$, where both 
time-arguments  are counted from the initial quench. This `logarithmic diffusion'  was  observed in 
Ref.~\cite{BoSi09} using particle track  data by Courtland and Weeks~\cite{Courtland03}
and is present  in the cluster model discussed in Ref.~\cite{Becker14}. However, tradition dictates that 
colloidal data be plotted versus the time-lag $\Delta t=t-t_{\rm w}$. Using this variable, one
easily obtains $\langle \Delta x^2(\Delta t)\rangle \propto \ln ( 1 + \Delta t/t_{\rm w})$. Hence,
as a function of $\Delta t$,  the MSD grows linearly as $\sim\Delta t / t_{\rm w}\ll1$ and  as 
$\sim- \ln t_{\rm w} + \ln \Delta t$ for  $\Delta t \gg t_{\rm w}$.  In Fig.~\ref{fig:tiledMSD}, we observe this  
behavior in  the particle tracking  data of  Ref.~\cite{Yunker09} at the highest area fractions, $84\%$   and $82\%$, deep in the jammed phase. The left-hand panel shows that the MSD grows 
initially weakly as $\sim\Delta t/t_{\rm w}$, then crosses over for  $\Delta t \gg t_{\rm w}$ to the
predicted logarithmic growth. The right-hand panel of Fig.~\ref{fig:tiledMSD} shows that the MSD data for 
different $t_{\rm w}$ and $t$ can, indeed, be collapsed by plotting them as functions of $t/t_{\rm w}$. 
As we show further below, the collapse with $t/t_{\rm w}$ remains valid at other jammed
packing fractions while the logarithmic slope of the plot increases with decreasing packing fraction. Only at 
densities too close to unjamming does the collapse begin to fail.
 
 \subsection{Persistence}
 \label{Persistence}
 \vspace{-0.25cm}
Not considered in Ref.~\cite{Yunker09} is the decay of  the persistence, i.e., the probability that a particle
does not experience irreversible motion between times $t_{\rm w}$ and $t$. This quantity gives a clear picture of 
the spatial heterogeneity of glassy
systems: If particle motion follows a regular Poisson process, the persistence decays exponentially 
in $\Delta t=t-t_{\rm w}$. In a log-Poisson process, this becomes an exponential in $\ln\left(t/t_{\rm w}\right)$,
leading to a power-law decay in $t/t_{\rm w}$. This prediction was already  borne out by simulations of jammed
Lennard-Jones systems~\cite{ElMasri10}, plotted as a function of lag-time $\Delta t$ for different waiting 
times $t_{\rm w}$. There, after an initial plateau for  $\Delta t \ll t_{\rm w}$, the data decays as a power 
law for $\Delta t \gg t_{\rm w}$. Analogous data from cluster-model simulations~\cite{Becker14} have 
reproduced this behavior from the record dynamics and demonstrated a successful collapse of the data as a 
function $t/t_{\rm w}$. 

\begin{figure}[t!]
\hfill{}\includegraphics[width=1\columnwidth]{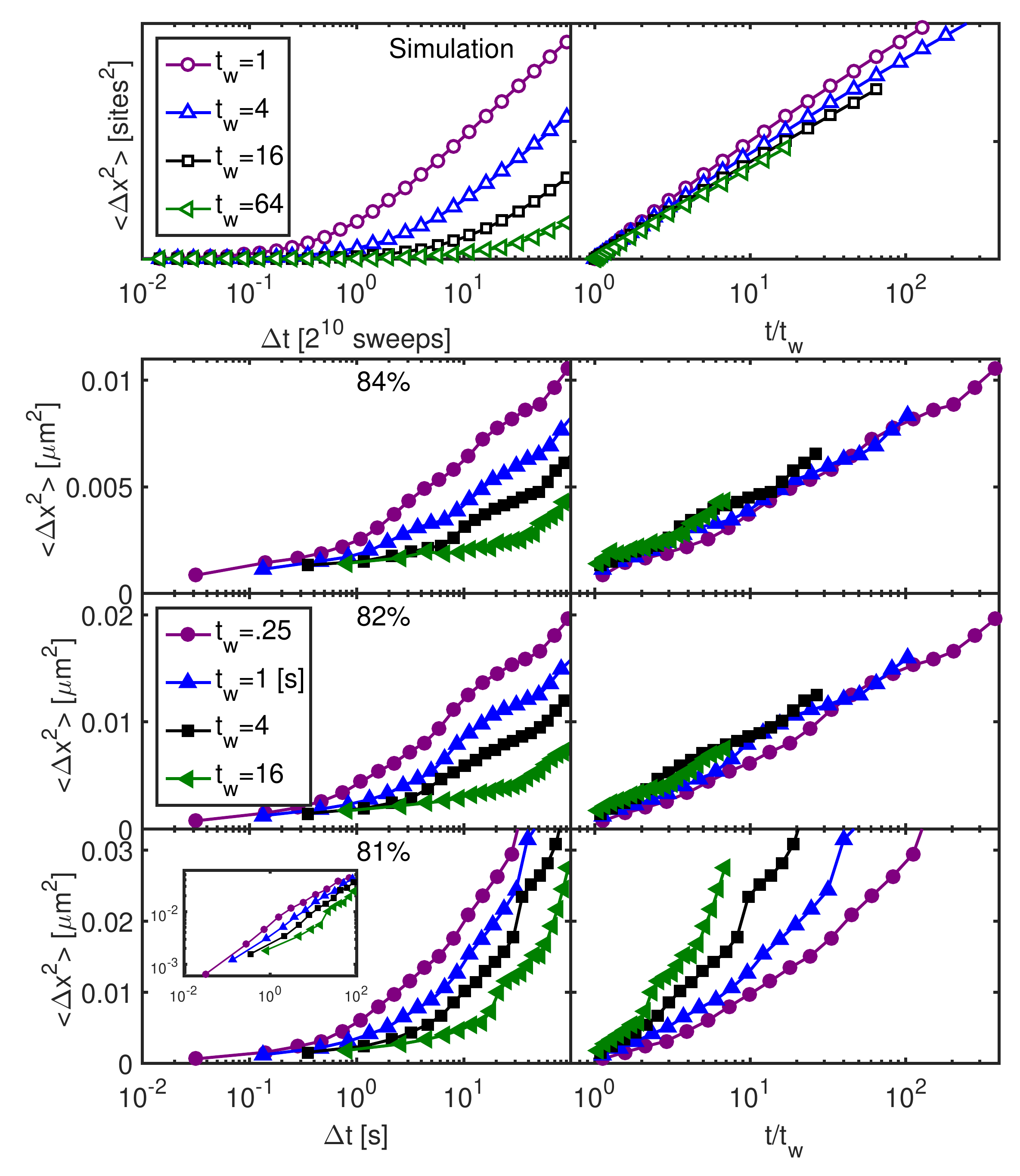}\hfill{}
 \vspace{-0.4cm}
\protect\caption{Mean square displacement in simulations and in the experiments~\cite{Yunker09} on aging 
colloids at several area fractions. Panels of the same row contain the exact same data, however, each panel in the left
column is plotted against the conventional lag-time  $\Delta t=t-t_{\rm w}$, and against the logarithmic 
time-ratio $\ln\left(t/t_{\rm w}\right)$ suggested by record dynamics in the right column. The top row shows the
simulation data, and the bottom three rows are experimental data arranged by decreasing area fractions, 
with $\approx84\%$, $\approx82\%$, and $\approx81\%$, from top to bottom.  Time units for simulation data are in multiples of $2^{10}$ Monte Carlo sweeps, 
to avoid transients after the quench in the early decades. The inset to the bottom left panel contains the same data as 
that panel, but plotted on a log-log scale.}
 \vspace{-0.3cm}
\label{fig:tiledMSD}
\end{figure}

\begin{figure}[t!]
\hfill{}\includegraphics[width=1\columnwidth]{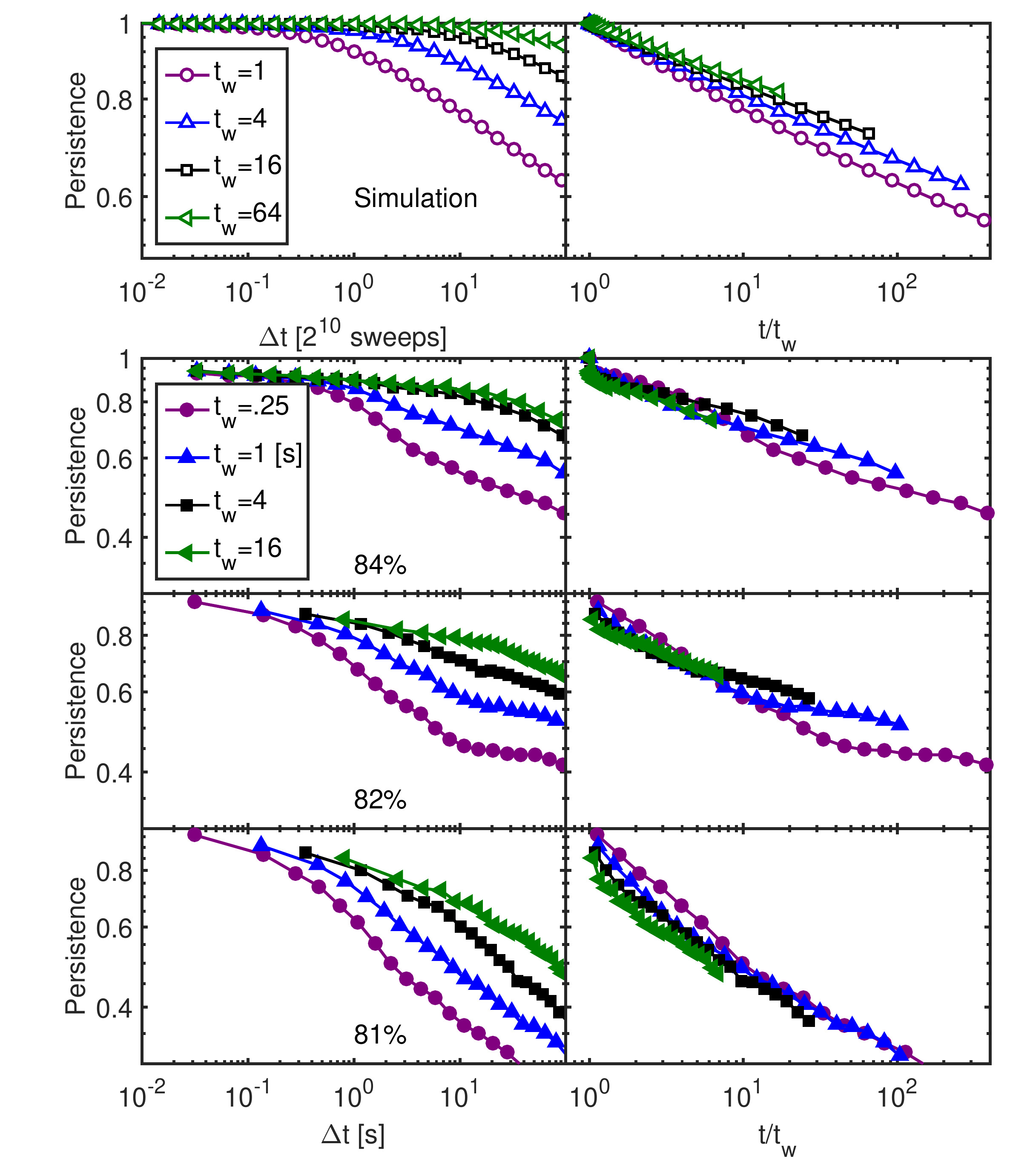}\hfill{}
 \vspace{-0.4cm}
\protect\caption{Persistence in simulations and in the experiments~\cite{Yunker09} on aging 
colloids at several area fractions. The arrangement of panels is identical to that of Fig.~\ref{fig:tiledMSD}.}
 \vspace{-0.3cm}
\label{fig:tiledPers}
\end{figure}

In the lower panels of Fig.~\ref{fig:tiledPers}, we show the experimental persistence data, plotted as functions 
of both, the lag time and $t/t_{\rm w}$, for several system ages $t_{\rm w}$.  Here, a threshold 
of $0.04\mu$m was used to separate reversible in-cage motion from a significant displacement after  $t_{\rm w}$. 
As for MSD in Fig.~\ref{fig:tiledMSD}, at the highest area fractions, $84\%$   and $82\%$, the data for the 
jammed system collapses on a $t/t_{\rm w}$-scale, but here in form of the predicted power-law. Overall, the
relaxation dynamics is  faster for less-jammed systems, such as that at $81\%$ area fraction shown in the lowest 
panels of Fig.~\ref{fig:tiledPers}. However, the persistence collapse does not break down as much at the packing 
fraction near unjamming when compared to the MSD. This might reflect the fact that supercooled systems already 
exhibit considerable heterogeneity~\cite{Weeks00}.

\subsection{Mobility Correlations}
\begin{figure}
\hfill{}\includegraphics[bb=0bp 0bp 890bp 400bp,clip,width=1\columnwidth]{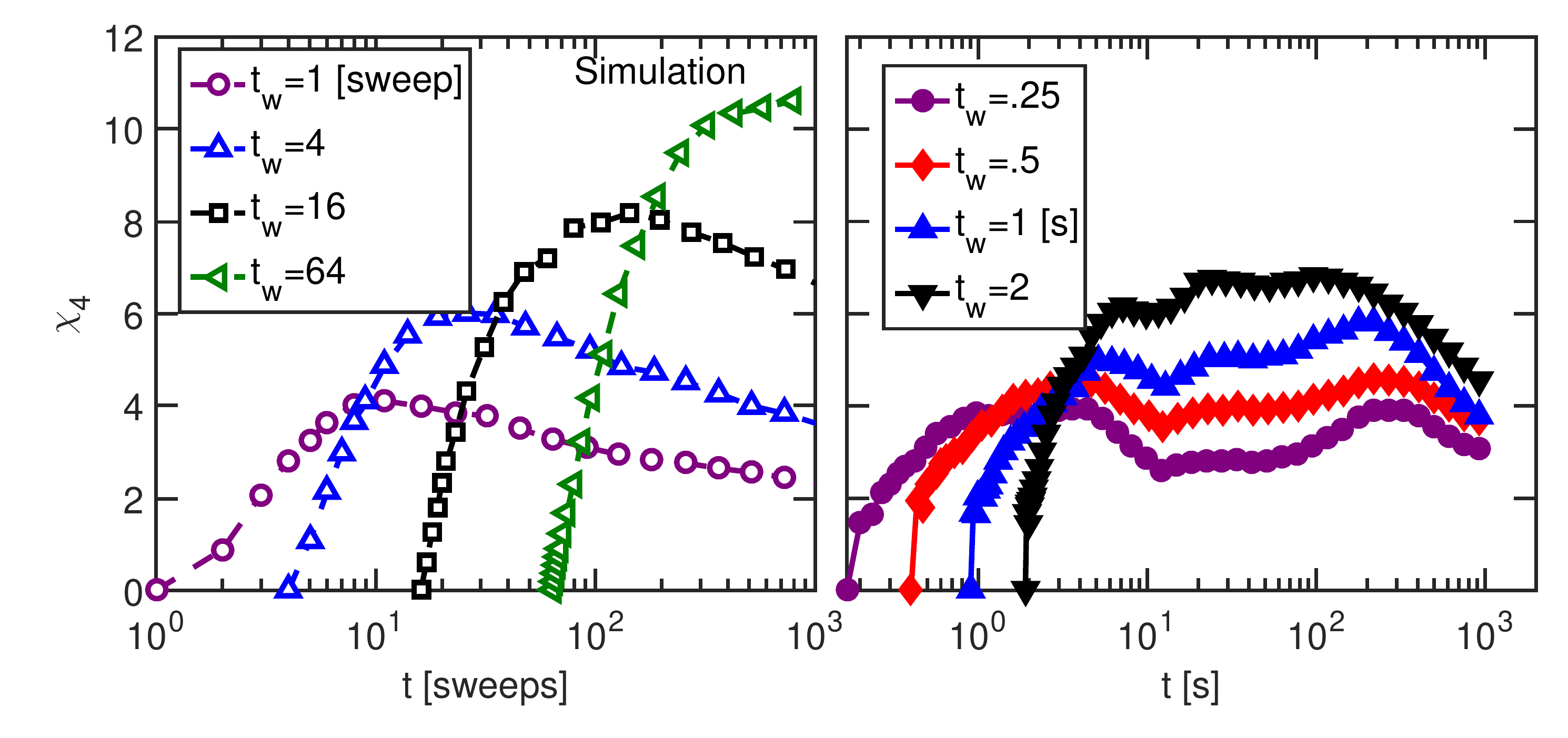}\hfill{}
 \vspace{-0.4cm}
\caption{\label{fig:x4}Growth of length scale of dynamic heterogeneity as
measured by the four-point susceptibility $\chi_{4}\left(t,t_{\rm w}\right)$,
which is a measure of the mobile particles at time $t$ that had been
mobile also at time $t_{\rm w}$. The height of each peak is proportional
to the number of particles involved in a significant rearrangement
event at time $t_{\rm w}$. Simulations of the cluster model ~\cite{Becker14}
(left) have shown that the size of those events grows $\sim\ln t_{\rm w}$,
reflected in the peak-height of $\chi_{4}$. The experimental
data (right) shows a discernible peak growing monotonically with $t_{\rm w}$ in an approximately
logarithmic manner  consistent with the record dynamics prediction.
 }
\vspace{-0.3cm}
\end{figure}

 \vspace{-0.25cm}
In Ref.~\cite{Yunker09}, the size of rearrangement events was characterized by the number of contiguous particles 
with unusually high speeds. While this approach is reasonable and common for detecting the presence of dynamic 
heterogeneity in tracking experiments, more precision is required to verify if the growth of event sizes with age 
corresponds to record dynamics. Therefore, the mobility-mobility correlation function $\chi_{4}$, widely used for 
supercooled liquids~\cite{Berthier05,DH2011,Berthier11}, has been proposed for this purpose~\cite{Parsaeian08,Maggi12,Becker14}; 
it is calculated as the ensemble variance of the total mobility at \emph{two} times: 
\begin{equation}
M(t,t_{\rm w})=\sum_{i}\exp(-|\vec{\mathbf{r}}_{i}(t)-\vec{\mathbf{r}}_{i}(t_{\rm w})|/d),
\label{eq:Mobility}
\end{equation}
where $d=0.04\mu$m is a typical length scale to distinguish mobile
from immobile particles, here taken to be the same to as the persistence
threshold above. The four-point correlation function $\chi_{4}$ is
then defined as
\begin{equation}
\chi_{4}(t,t_{\rm w})=\left\langle M(t,t_{\rm w})^2\right\rangle -\left\langle M(t,t_{\rm w})\right\rangle^2.
\label{eq:DefChi4}
\end{equation}
It is an as-of-yet unconfirmed prediction of record dynamics that
the peak of $\chi_{4}$ grows as $\log t_{\rm w}$~\cite{Becker14}. 
Reasoning by analogy with the cluster model, 
particles that move at  time $t_{\rm w}$ belong to collapsing  clusters of
average size $\sim\log t_{\rm w}$. The re-activation of
those particles requires that a cluster of the same size is reformed and collapses. Such process 
requires a  time-interval $\Delta t_{{\rm peak}}$ beyond $t_{\rm w}$ that grows monotonically 
with $t_{\rm w}$: For times much shorter than $\Delta t_{{\rm peak}}$, too few particles have
a chance of reactivation, while for  times much longer  than $\Delta t_{{\rm peak}}$
most of them already did re-activate and, hence, de-correlate from their mobility at $t_{\rm w}$.
For the experimental data, consisting of single runs, there is no
suitable ensemble to average over, so $\chi_{4}$ is approximated here
by dividing a sample into four quadrants and taking the variance across
those. This measure is plotted in Fig.~\ref{fig:x4}. There are large
fluctuations in $\chi_{4}$ due to the unavoidable interdependence
of the regions within a sample and the lack of statistics. In spite
of this, the curves for different waiting times show regularly increasing
peak heights for exponentially increasing waiting times, as predicted
in simulations of record dynamics~\cite{Becker14}, also shown in Fig.~\ref{fig:x4}.

\subsection{Data at lower Area Fraction}
\label{Area_Fraction}
 \vspace{-0.25cm}
The  data from Ref.~\cite{Yunker09} can be used to test the robustness of 
record dynamics over a range of area fractions. We have stated that specifics of a system do not impact the 
schematic description of the free-energy landscape in Fig.~\ref{fig:FLandscape}, as long as the system is jammed. 
Indeed, the experimental data for the fully jammed systems at the two area fractions $\approx84\%$ 
and $\approx82\%$  in Figs.~\ref{fig:tiledMSD}-\ref{fig:tiledPers} both show identical behavior for MSD as well 
as for persistence but for a different unit of time. Thus, only the slopes  for the respective collapse of the data is 
affected (note the scale on each y-axis), indicating less MSD and more persistence within a fixed amount of 
wall-time for denser systems. 

The bottom panels of the same figures refer to a system at an area fraction
of $\approx81\%$, whose  behavior is consistent with being unjammed or being too near to the jamming 
transition. Thus, record dynamics is not expected to apply. In fact, the inset of the bottom left panel 
of Fig.~\ref{fig:tiledMSD} shows that the MSD  
exhibits nearly linear scaling vs. $\Delta t$, which is nearly diffusive behavior. 
Due to a  weak $t_{\rm w}$ dependence, the MSD data do not    collapse 
when plotted against $\ln\left(t/t_{\rm w}\right)$ in the  rightmost panel. However, 
 the data collapse in the bottom panels of Fig.~\ref{fig:tiledPers} suggests that 
 a semblance of dynamic heterogeneity is retained near the jamming 
transition in the persistence data. 

\section{Comparison with Simulations of the Cluster Model}
 \vspace{-0.25cm}
Becker et al~\cite{Becker14} simulated aging over 15 decades using a real-space implementation of 
record dynamics called the 'cluster model'. In this model, particles fill a lattice where in-cage rattle is explicitly 
course-grained away. All particles are 
assigned to contiguous clusters. There is a chance for a cluster to 'break up', which is exponentially unlikely in the 
size of the cluster. When this happens the cluster's constituent particles move minutely, and neighboring clusters 
spread into the broken cluster's territory. This process yields a breakup event rate proportional to $1/t$ and an 
average cluster size proportional to  $\ln t$. The $1/t$ event rate is observed in 
the experimental data in Fig.~\ref{fig:Decay}. The growth of cluster size is a prediction of record dynamics that is 
consistent with observations of cluster sizes of fast particle in Fig.~2b in Ref.~\cite{Yunker09}.

Here, the mean square displacement and persistence curves obtained from simulations are restricted to 3 
decades of scaling to match the range of the experiments. When plotted against lag-time, shown in the upper left 
panel of Fig.~\ref{fig:tiledMSD}, the simulation MSD  exhibits the commonly expected plateau 
for $\Delta t \ll t_{\rm w}$ and rise for  $\Delta t \gg t_{\rm w}$. Averaged over so many runs that error bars 
become invisible, the simulation curves clearly demonstrate the logarithmic 
behavior expected from record dynamics. The plateaus in the experimental data at high area fractions in Fig.~\ref{fig:tiledMSD} have a slight 
slope $\sim\Delta t/t_{\rm w}\ll1$ due to the in cage motion. Both the simulation and experimental data collapse 
to a uniform logarithmic growth at all $t_{\rm w}$ when plotted against $t/t_{\rm w}$.

Persistence in the simulation is 
calculated as the fraction of particles that have  not moved to a different lattice site. As with the MSD, due to the lack of 
in-cage rattle, the systems appear frozen for lag-times shorter 
than the waiting time. 
The persistence data from simulations shown in the top row of Fig.~\ref{fig:tiledPers} demonstrate the power-law 
decay predicted by record dynamics. For the experimental data, a suitable metric had to be determined to classify 
particles as persistent. It was found that setting a distance threshold and taking the fraction of particles that move 
beyond that distance from their position at $t_{\rm w}$, robustly yields results qualitatively similar to the 
simulation curves.  The experimental persistence curves shown here are all produced using a threshold 
of 0.04$\mu$m.

\section{Conclusions}
 \vspace{-0.25cm}
Our detailed analysis of the experimental particle tracking data of Yunker et al.~\cite{Yunker09}
 \emph{i)} Shows that the rate of irreversible particle rearrangements falls off as the inverse of the 
system age;  \emph{ii)} Confirms~\cite{BoSi09} that the  positional  variance due to particle motion 
between $t_{\rm w}$ and $t$ grows as $\ln t/t_{\rm w}$, a diffusive behavior in the logarithm of 
time;  \emph{iii)} Shows that persistence data can be scaled in the same way, as function of $t/t_{\rm w}$; 
\emph{iv)} Shows that the peak of the four-point susceptibility $\chi_{4}\left(t,t_{\rm w}\right)$
grows in a way consistent with $\ln t_{\rm w}$;
\emph{v)} Confirms  that the range of applicability of record dynamics is, as predicted, limited
to sufficiently dense colloidal systems. Finally, the analysis of the experimental data 
concurs simulational data from   our `cluster model'~\cite{BoSi09,Becker14}.

The agreement between  experimental data  and record dynamics predictions  for all packing fractions in the jammed regime (and only there)  
confirms that aging dynamics is controlled in each instance by a small set of active variables,  
which move intermittently in time. That such variables cluster in space is well known as spatial heterogeneity
and can be seen in the data of Ref.~\cite{Yunker09}. Even though more accurate experimental measurement 
of $\chi_{4}$ are required to ascertain whether   the  clusters grow  logarithmically in  time,
there is clear experimental evidence  that they do grow. 

Continuous Time Random Walks (CTRW) are  
most widely used to coarse-grain the dynamics of   systems jumping from trap to trap via rare intermittent events, see e.g.~\cite{Heuer08}.
Confusingly, CTRW and RD predictions are partly overlapping, while their physical mechanisms  are very different.
Importantly,
as one of us  argued in Ref.~\cite{Sibani13},  CTRW are 
for a number of reasons inadequate descriptions of  aging. 
Here we  can only  add  that the emergence of a growing physical length scale 
   suffices to rule out a  renewal process as the  underlying mechanism for aging in dense colloids.
Real-space mesoscopic objects whose  lifetime  grows exponentially with their size or, equivalently,
whose  characteristic length  grows logarithmically in time,  are arguably a key feature of off-equilibrium
glassy dynamics and deserve further experimental investigation.

\acknowledgments
The authors are indebted to the V. Kann Rasmussen Foundation for support.
SB and DMR thank SDU for its hospitality. SB and DMR are further supported
by the NSF through grant DMR-1207431. PY thanks A.G.~Yodh, K.B.~Aptowicz, and Z.~Zhang for useful conversations relating to Ref.~\cite{Yunker09}.

\end{document}